\documentclass[10pt,a4]{article}

\usepackage{latexsym,amsmath,amscd,amssymb,graphicx}

\begin{document}
\begin{titlepage}

\title{A new view on relativity:\\Part 1. Kinematic relations between inertial and
relativistically accelerated systems based on symmetry}

\author{Yaakov Friedman\\Jerusalem College of
Technology\\ P.O.B. 16031 Jerusalem 91160\\
Israel\\e-mail:friedman@jct.ac.il}

\date{}
\maketitle

\begin{abstract}

Several new ideas related to Special and General Relativity are
proposed.  The black-box method is  used for the synchronization
of the clocks and the space axes between two inertial systems or
two accelerated systems and for the derivation of the
transformations between them. There are two consistent ways of
defining the inputs and outputs to describe the transformations
and relative motion between the systems. The standard approach
uses a mixture of the two ways. By formulating the principle of
special and general relativity as a symmetry principle we are able
to specify these transformations to depend only on a constant.

The transformations become Galilean if the constant is zero.
Validity of the Clock Hypothesis for uniformly accelerated systems
implies zero constant. If the constant is not zero, we can
introduce a metric under which the transformations become self
adjoint.  In case of inertial systems, the metric is the Minkowski
metric and we obtain a unique invariant maximal velocity. The ball
of the relativistically admissible velocities is a bounded
symmetric domain under projective maps.  For uniformly accelerated
systems the existence of an invariant \textit{maximal
acceleration} is predicted. This is the only method of describing
transformations between uniformly accelerated systems without
assuming the Clock Hypothesis.


\end{abstract}

\end{titlepage}

\section{Introduction}

In this article we present the first steps of a program for a
unifying language for physics based on the the theory of bounded
symmetric domains. Most results presented here can be found with
details in \cite{F04} and \cite{FG5}.

In this article we will derive the transformations between
inertial and uniformly accelerated systems based only on the
principles of special and general relativity, respectively, and
the symmetry implied by them. We use the black-box method for the
synchronization of the clocks and the space axes between the two
systems and for the derivation of the transformations between
them. We will use two consistent ways of defining the inputs and
outputs for description of the the transformations and relative
motion between the systems. In each way we will specify the form
of the transformation and the  meaning of its components. By
formulating the principles of special and general relativity as
symmetry principles we are able to explicitly  specify these
transformations (with dependence only on a constant).

The transformations become Galilean if the above constant is zero.
Validity of the Clock Hypothesis implies that the constant is zero
for uniformly accelerated systems. If the constant is not zero, in
case of inertial systems  we obtain a unique invariant maximal
velocity and for uniformly accelerated systems the existence of an
invariant maximal acceleration is predicted.

In forthcoming Part 2 we will discuss new ideas of relativistic
evolution and its connection with the invariance of the ball of
relativistically admissible velocities or accelerations. Such
evolution is defined by the Lie algebra for the automorphism
groups of these balls and via a new Equivalence Principle. We will
explain how a new dynamic variable, which is related to the
spinors, helps to solve relativistic dynamic equations.

\section{Transformations between inertial systems based on the principle of relativity}

In this section, we derive the space-time transformation between
two inertial systems, using only the isotropy of space and
symmetry, which follow from the principle of relativity. The
transformation will be defined uniquely except for a constant $e$,
which depends only on the process of synchronization of clocks
inside each system.

 We begin with two inertial systems. Typically, there are
events which are observable from both systems. We will assume that
each system has a way of defining space and time parameters
describing an observed event.

Newton's First Law states that an object moves with constant
velocity in an inertial system if there are no forces acting on it
or if the sum of all forces on it is zero. Such a motion is called
free motion and is described by straight lines in the space-time
continuum. Free motion in one inertial system will be observed as
free motion in {any} inertial system. This means that the
space-time transformations will map lines to lines. Thus, if we
chose space and time as the parameters with which to describe
events in the inertial systems and common space axes at time
$t=0$, the space-time transformation will be a \textit{linear}
transformation.

  An event observed by two inertial systems can be
considered as a ``two-port linear black box". Each side of the box
correspond to an inertial system. One port on each side defines
the time of the event, while the second one defines the space
coordinates of the event. The observation of the event by each
system provides information on the transformation between these
systems. Since our intuition does not work well in a dynamically
varying environment and in cases of extremely high velocities, the
black box approach is preferable to the approach which assumes
\textit{a priori} some properties of the transformation.

\subsection{ Identification of symmetry inherent in the principle of special relativity}

Albert Einstein formulated  the
 \textit{principle of special
relativity} (\cite{E55}, p.25): ``If $K$ is an inertial system,
then every other system $K'$ which moves uniformly and without
rotation relatively to $K$, is also an inertial system; the laws
of nature are in concordance for all inertial systems." By the
principle of special relativity, the space-time transformation
between the systems will depend only on the choice of the space
axes, the measuring devices (consisting of rods and clocks) and
the relative motion between these systems.

 The relative motion
between two inertial systems is described by their
\index{velocity!relative} \textit{relative velocity}. We denote by
$\mathbf{b}$ the relative velocity (called the boost) of $K'$ with
respect to $K$ and by $\mathbf{b}'$ the relative velocity of $K$
with respect to $K'$. If we choose the measuring devices in each
system to be the same and choose the axes in such a way that the
coordinates of $\mathbf{b}$ are equal to the coordinates of
$\mathbf{b}'$, then the space-time transformation $S$ from $K$ to
$K'$ will be equal to the space-time transformation $S'$ from $K'$
to $K$. Since, in general, $S'=S^{-1}$, in this case we will have
$S^2 =I$. Such an operator $S$ is called a
\index{transformation!symmetry} \index{symmetry!operator}
\textit{symmetry}. Thus, the principle of special relativity
implies that with an appropriate choice of axes and measuring
devices, the space-time transformation $S$ between two inertial
systems is a symmetry.

 \subsection{Synchronization of the clocks and the space axes in two
inertial systems.}\label{synchron}

As mentioned above, in order for the space-time transformation to
be a symmetry we have to choose the measuring devices in each
system to be the same and choose the axes in such a way that the
coordinates of $\mathbf{b}$ are equal to the coordinates of
$\mathbf{b}'$. To do this, we will synchronize the two systems by
observing events from each system and comparing the results.
System 1 begins with the following configuration. There is a set
of three mutually orthogonal space axes and a system of rods. In
this way, each point in space is associated with a unique vector
in $R^3$. In addition, there is a clock at each point in space,
and all of the clocks are synchronized to each other by some
synchronization procedure. System 2 has the same setup, only we do
not assume that the rods of system 1 are identical to the rods of
system 2, nor do we assume that the clock synchronization
procedure in system 2 is the same as that of system 1.

First, we synchronize the origins of the frames.  Produce an event
$E_0$ at the origin $O$ of system 1 at time $t=0$ on the clock
positioned in system 1 at $O$. This event is observed at some
point $O'$ in system 2, and the system 2 clock at $O'$ shows some
value $t'=t'_0$. Translate the origin of system 2 to the point
$O'$ (without rotating). Subtract $t'_0$ from the system 2 clock
at $O'$. Synchronize all of the system 2 clocks to this clock.
This completes the synchronization of the origins.

Next, we will adjust the $x$-axis of each system. Since the
systems are varying dynamically, any adjustment in a system could
be done only by use of objects which are static  in this system.
Note that system 2 is moving with some (perhaps unknown) constant
velocity $\mathbf{b}$  with respect to system 1 and that the
origin $O'$ of system 2 was at the point $O$ of system 1 at time
$t=0$. Therefore, the point $O'$ will always be on the static line
$\mathbf{b}t$ in system 1. Rotate the axes in system 1 so that the
new negative $x$-axis coincides with the ray
$\{\mathbf{b}t:t>0\}$. Similarly, system 1 is moving with some
constant velocity $\mathbf{b'}$ with respect to system 2, and the
origin $O$ of system 1 was at the point $O'$ of system 2 at time
$t'=0$. Therefore, the point $O$ will always be on the line
$\mathbf{b'}t$ in system 2. Rotate the axes in system 2 so that
the new negative $x'$-axis coincides with the ray
$\{\mathbf{b'}t:t>0\}$. The two $x$-axes now coincide as lines and
point in opposite directions. We are finished manipulating the
axes and clocks of system 1 and will henceforth refer to system 1
as the inertial frame $K$.  However, it still remains to
manipulate system 2, as we must adjust the $y'$- and $z'$-axes of
system 2 to be parallel and oppositely oriented to the
corresponding axes of $K$.

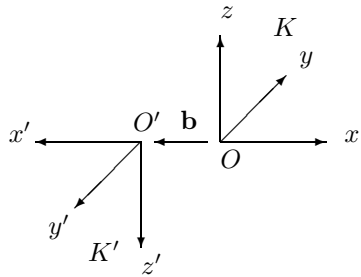
\begin{figure}[h!]
  \centerline{
   \begin{picture}(200,120)(0,0)
   \put (100,50){\vector(1,0){40}}
   \put (100,50){\vector(0,1){40}}
   \put (100,50){\vector(1,1){25}}
   \put (70,50){\vector(-1,0){40}}
   \put (70,50){\vector(0,-1){40}}
   \put (70,50){\vector(-1,-1){25}}
   \put (95,50){\vector(-1,0){20}}
   \put (147,50){$x$}
   \put (20,50){$x'$}
   \put (100,97){$z$}
   \put (70,00){$z'$}
   \put (130,80){$y$}
   \put (35,15){$y'$}
   \put (85,55){$\mathbf{b}$}
   \put (120,90){$K$}
   \put (50,5){$K'$}
   \put (100,40){$O$}
   \put (67,54){$O'$}
   \end{picture}}
  \caption{Two symmetric space reference frames. The relative velocity
  of the inertial system $K'$ with respect to $K$ is $\mathbf{b}$. The coordinates
  of $\mathbf{b}$ in $K$ are equal to the coordinates (in $K'$) of the relative velocity
  of the system $K$ with respect to $K'$.}\label{Frefframe}
 \end{figure}

To adjust the $y'$-axis of system 2, produce an event $E_1$ at the
point $\mathbf{r}=(0,1,0)$ of $K$. This event is observed in
system 2 at some point $\mathbf{r}'$. Rotate the space axes of
system 2 around the $x'$-axis so that $\mathbf{r}'$ will lie in
the new $x'$-$y'$ plane and have a negative $y'$ coordinate
$y_{1}'$. Change the space scale to make this new coordinate
$y_{1}'$ equal -1. After this rotation, the $z$-axis of $K$ and
the $z'$-axis of system 2 will be parallel. We need to make sure
that they have opposite orientations. Produce an event $E_2$ at
the point $\mathbf{r}=(0,0,1)$ of $K$. This event is observed in
system 2 at some point $\mathbf{r}'$. If the $z'$ coordinate of
$\mathbf{r}'$ is positive, reverse the direction of the $z'$-axis.
This completes the adjustment of the space axes of the two
systems. See Figure \ref{Frefframe}.

We will call such a frame a \textit{symmetric frame}. Finally,
change the scale of the time in system 2 to make the relative
velocity of $K$ with respect to $K'$ be equal to $\mathbf{b}$, the
relative velocity of $K'$ with respect to $K$.

 \subsection{Choice of inputs and outputs}

There are two ways to define the inputs and outputs for such a
transformation.

\subsubsection{Cascade connection}

\begin{figure}[h!]
\centering
 \scalebox{0.35}{\includegraphics{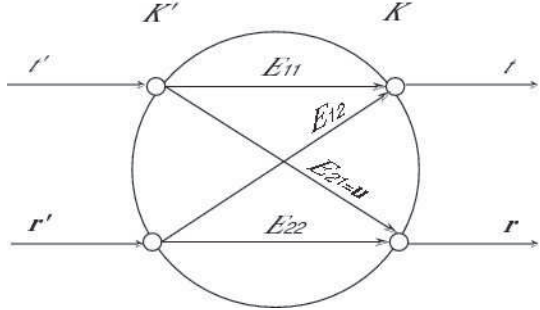}}
  \caption[The cascade connection for space-time transformations.]{The cascade connection for space-time transformations. The
  circle represents a black box.
   One side has two input ports: the time $t'$ and the
  space  $\mathbf{r}'$ coordinates of an event in system $K'$. The
  other side has two output ports:  the time $t$ and the
  space  $\mathbf{r}$ coordinates of the same event in system $K$.
  The linear operators $E_{ij}$ represent the functional
  connections between the corresponding ports.}
  \label{cascade}
\end{figure}

 The first one, called the \index{cascade connection}
 \index{connection!cascade} \textit{cascade
connection}, takes time and space of one of the systems, say $\left(  \begin{array}{r} t'\\
 \mathbf{r}' \end{array} \right)$ of $K'$, as
input, and gives time and space of the
second system, say $\left(  \begin{array}{r} t\\
 \mathbf{r} \end{array} \right)$ of $K$, as output (see Figure
 \ref{cascade}).
 Note that we use a circle instead of the usual box to represent a black-box. This
 is done in order that the connection between any
  two ports will be displayed  \textit{inside} the box (see Figure
 \ref{Fhybrid}).

The cascade connection is the one usually used in special
relativity.

We represent the linear transformation induced by the cascade
connection by a $ 4\times 4$ matrix $E$, which we decompose into
four block matrix components $E_{ij}$, as follows:
\begin{equation}\label{comp1}
 \left( \begin{array}{c}  t\\ \mathbf{r}
          \end{array} \right)=E\left( \begin{array}{c}  t'\\
          \mathbf{r}'
          \end{array} \right)= \left(
         \begin{array}{rr}
              E_{11} & E_{12} \\
              E_{21}& E_{22}
          \end{array} \right)
          \left( \begin{array}{r}  t'\\ \mathbf{r}'
          \end{array} \right).
\end{equation}
To understand the meaning of the blocks, assume that the system
$K'$ is an airplane. Let $t'$ be the time between two events (say
crossing two lighthouses) measured by a clock at rest at
$\mathbf{r}'=0$ on the airplane. The time difference $t$ of the
same two events measured by synchronized clocks at the two
lighthouses (in system $K$, the earth) will be equal to
$t=E_{11}t'$. If we denote the distance between the lighthouses by
$\mathbf{r}$, then $\mathbf{r}=E_{21}t'$, and $E_{21}$ is the
so-called \textit{proper velocity}\index{proper velocity}
\label{propvel} of the plane. Generally, the proper velocity
$\mathbf{u}$ of an object (the airplane) in an inertial system is
the ratio of the space displacement $d\mathbf{r}$ in lab system
(the earth) divided by the time interval, called the proper time
interval $d\tau$, measured by the clock moving with the object (on
the plane). Thus,
\begin{equation}\label{propervelocity}
  \mathbf{u}=\frac{d\mathbf{r}}{d\tau}.
\end{equation}

\subsubsection{Hybrid connection}\label{hybridsec}

\begin{figure}[h!]
  \centering
  \scalebox{0.35}{\includegraphics{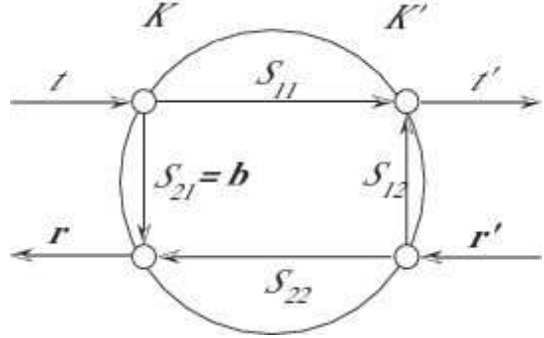}}
  \caption[The hybrid connection for space-time transformations.]{The hybrid connection for space-time transformations. The
  circle represents a black box.
   The two input ports are the time $t$ of an event, as measured
 in system $K$, and its space  coordinates $\mathbf{r}'$,  as measured
  in system $K'$. The two output ports are
  the time $t'$ of the same event, calculated in system $K'$, and its
  space  $\mathbf{r}$ coordinates, calculated in
  $K$. The linear operators $S_{ij}$ represent the functional
  connections between the corresponding ports.
  For instance, to define the map $S_{21}$, we consider
an event that occurs at $O'$, corresponding to input
$\mathbf{r}'=0$, at time $t$ in $K$. Then $S_{21}t$ represent the
space displacement of $O'$ in $K$ during time $t$, which is, by
the definition, the relative velocity $\mathbf{b}$ of system $K'$
with respect to system $K$.}\label{Fhybrid}
\end{figure}

The second type of connection, called the
\index{connection!hybrid} \index{hybrid connection} \textit{hybrid
connection}, uses time of one of the systems, say $t$ of $K$, and
the space coordinates
$\mathbf{r}'$ of the second system $K'$, as input, and gives $\left(  \begin{array}{r} t'\\
 \mathbf{r} \end{array} \right)$
as output (see Figure \ref{Fhybrid}).
 Usually we use relative \textit{velocity} (not relative proper velocity) to describe the relative
position between inertial systems. To define the relative position
of system $K'$ with respect to $K$,  we consider an event that
occurs at $O'$, corresponding to $\mathbf{r}'=0$, at time $t$, and
express its position $\mathbf{r}$ in $K$.  If we denote by
$\mathbf{b}$ the uniform velocity of system $K'$ with respect to
$K$, then
\begin{equation}\label{g}
  \mathbf{r}=\mathbf{b}t.
\end{equation}
Note our use of the hybrid connection. In this section we will use
the hybrid connection in order to keep the relative velocity as
the description of relative position between the systems.
Furthermore, a bounded symmetric domain is obtained only for the
velocities and not for the proper velocities.

We denote by $S_{\mathbf{b}}$ the space-time transformation, using
the hybrid connection, for two inertial systems with relative
velocity $\mathbf{b}$ between them. Thus, for the transformation
$S_{\mathbf{b}}$, we choose the inputs to be the scalar $t$, the
time of the event in $K$, and the three-dimensional vector
$\mathbf{r}'$ describing the position of the event in $K'$.  Then
our outputs are the scalar $t'$, the time of the event in $K'$,
and the three-dimensional vector $\mathbf{r}$ describing the
position of the event in $K$. As above with respect to the cascade
connection, here we also decompose the $4 \times 4$ matrix
$S_{\mathbf{b}}$ into block components:
\begin{equation}\label{comp}
 \left( \begin{array}{c}  t'\\ \mathbf{r}
          \end{array} \right)=S_{\mathbf{b}}\left( \begin{array}{c}  t\\ \mathbf{r}'
          \end{array} \right)= \left(
         \begin{array}{rr}
              S_{11} & S_{12} \\
              S_{21}& S_{22}
          \end{array} \right)
          \left( \begin{array}{r}  t\\ \mathbf{r}'
          \end{array} \right).
\end{equation}
(see Figure \ref{Fhybrid}).

We explain now the meaning of the four linear maps $S_{ij}$. To
define the maps $S_{21}$ and $S_{11}$, consider an event that
occurs at $O'$, corresponding to $\mathbf{r}'=0$, at time $t$ in
$K$. Then $S_{21}(t)$ expresses the position of this event in $K$,
and $S_{11}(t)$ expresses the time of this event in $K'$.
Obviously, $S_{21}$ describes the relative velocity of $K'$ with
respect to $K$, and
\begin{equation}\label{omega}
  S_{21}(t)=\mathbf{b}t,
\end{equation}
while $S_{11}(t)$ is the time shown by the clock positioned at
$O'$ of an event occurring at $O'$ at time $t$ in $K$ and is given
by
\begin{equation}\label{teta}
  S_{11}(t)=\alpha t
\end{equation}
for some constant $\alpha$.

\begin{figure}[h!]
\centering
 \scalebox{0.3}{\includegraphics{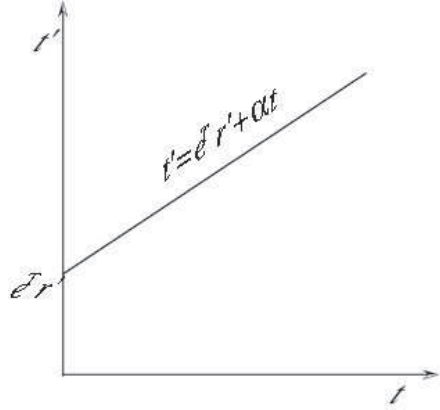}}
  \caption[The time of an event in two inertial systems]{The times $t'$ and $t$ of an event at space point $\mathbf{r}'$
  in system $K'$. The difference in timings is caused both by
  the difference in the rates of clocks (time slowdown) in each
  system and by the different synchronization of the clocks positioned
  at different space points.}
  \label{nonsynchron}
\end{figure}

To define the maps $S_{12}$ and $S_{22}$, we will consider an
event occurring  at time $t=0$ in $K$ in space position $
\mathbf{r}'$ in $K'$.  Then $S_{12} (\mathbf{r}')$ will be the
time of this event in $K'$, and $S_{22} (\mathbf{r}')$ will be the
position of this event in $K$.  Note that $S_{12} (\mathbf{r}')$
is also the time difference of two clocks, both positioned at time
$t=0$ at $\mathbf{r}'$ in $K'$, where the first one was
synchronized to the clock at the common origin of the two systems
within the frame $K'$, and the second one was synchronized to the
clock at the origin within the frame $K$. Thus $S_{12}$ describes
the \textit{non-simultaneity} \index{non-simultaneity}in $K'$ of
simultaneous events in $K$ with respect to their space
displacement in $K'$, following from the difference in
synchronization of clocks \index{clock
synchronization}\index{synchronization} in $K$ and $K'$. Since
$S_{12}$ is a linear map from $R^3$ to $R$, it is given by:
\begin{equation}\label{phi}
  S_{12}(\mathbf{r}')=
  {\mathbf{e}^T}\mathbf{r}',
\end{equation}
for some vector $\mathbf{e}\in R^3$, where $\mathbf{e}^T$ denotes
the transpose of $\mathbf{e}$. Note that
${\mathbf{e}^T}\mathbf{r}'$ is the dot product of ${\mathbf{e}}$
and $\mathbf{r}'$. See Figure \ref{nonsynchron} for the connection
between the time of events in two inertial systems.

 Finally, the map $S_{22}$ describes the transformation of the
 space displacement in $K$ of simultaneous events in $K$ with respect
 to their space displacement in $K'$, and it is given by
\begin{equation}\label{gamma}
  S_{22} (\mathbf{r}')=A\mathbf{r}'
\end{equation}
for some $3\times 3$ matrix $A$.

Note that the usual approach uses the cascade connection for the
space-time transformation and the relative velocity, which comes
from the hybrid connection, for the description of the relative
motion between the systems. However, space contraction, time
contraction and the measure of non-simultaneity are defined by use
of the hybrid connection.

\subsubsection{The transformation between cascade and hybrid connections}

Note that the matrices $E$ and $S_{\mathbf{b}}$ describing the
space-time transformations between two inertial systems using the
cascade and hybrid connections, respectively, are related by some
transformation $\Psi$. The transformation is defined by
\begin{equation}\label{EtoS1}
          \Psi \left(
         \begin{array}{rr}
              E_{11} & E_{12} \\
              E_{21}& E_{22}
          \end{array} \right)=\left(
\begin{array}{cc}
              E_{11}-E_{12}E_{22}^{-1}E_{21} & E_{12}E_{22}^{-1} \\
              -E_{22}^{-1}E_{21}& E_{22}^{-1}
          \end{array} \right).
\end{equation}
This transformation is called the \textit{Potapov-Ginzburg
transformation}. It can be shown that
\begin{equation}\label{EtoS}
 \left(
         \begin{array}{rr}
              S_{11} & S_{12} \\
              S_{21}& S_{22}
          \end{array} \right)=\Psi \left(
         \begin{array}{rr}
              E_{11} & E_{12} \\
              E_{21}& E_{22}
          \end{array} \right)
\end{equation}
and
\begin{equation}\label{StoE}
 \left(
         \begin{array}{rr}
              E_{11} & E_{12} \\
              E_{21}& E_{22}
          \end{array} \right)=\Psi \left(
         \begin{array}{rr}
              S_{11} & S_{12} \\
              S_{21}& S_{22}
          \end{array} \right).
\end{equation}
It is easy to check that $S_{\mathbf{b}}$ is a symmetry (that is,
$S_{\mathbf{b}}^2=I$) if and only if $E=\Psi(S_{\mathbf{b}})$ is a
symmetry.

\subsection{ Derivation of the explicit form of the
symmetry operator}

Our black box transformation can now be described by a $4\times4$
matrix $S_{\mathbf{b}}$ with block matrix entries  from
(\ref{omega}), (\ref{teta}), (\ref{phi}) and (\ref{gamma}):
\begin{equation}\label{bbt}
\left( \begin{array}{c}  t'\\ \mathbf{r}
          \end{array} \right)=S_{\mathbf{b}}\left( \begin{array}{c}  t\\ \mathbf{r}'
          \end{array} \right)=
          \left(
         \begin{array}{rr}
              \alpha & \mathbf{e}^T \\
              \mathbf{b}&  A
          \end{array} \right)
          \left( \begin{array}{r}  t\\ \mathbf{r}'
          \end{array} \right)
          . \end{equation}
If we now interchange the roles of systems $K$ and $K'$, we will
get a matrix $S'_{\mathbf{b}}$:

\begin{equation}\label{bbt1}
\left( \begin{array}{c}  t\\ \mathbf{r}'
          \end{array} \right)=S'_{\mathbf{b}}\left( \begin{array}{c}  t'\\ \mathbf{r}
          \end{array} \right)=
          \left(
         \begin{array}{rr}
              \alpha ' & {\mathbf{e'}^T} \\
              \mathbf{b}'&  A'
          \end{array} \right)
          \left( \begin{array}{r}  t'\\ \mathbf{r}
          \end{array} \right).\end{equation}
But the principle of relativity implies that switching the roles
of $K$ and $K'$ is nonrecognizable. Hence
\[\alpha=\alpha',\;\;\mathbf{e}^T={\mathbf{e'}^T},\;\;\mathbf{b}=
\mathbf{b}', \;\; A=A' .\]

 By combining (\ref{bbt}) and (\ref{bbt1}), we get $S_{\mathbf{b}}^2=I$,
 implying that $S_{\mathbf{b}}$ is a symmetry operator. Hence,
 \begin{equation}\label{sym}
\left(         \begin{array}{cc}
              \alpha & \mathbf{e}^T \\
              \mathbf{b}&  A
          \end{array} \right)
          \left( \begin{array}{rr}
           \alpha & \mathbf{e}^T \\
              \mathbf{b}&  A
          \end{array} \right)=\left( \begin{array}{rr}
           1 & \mathbf{0}^T \\
              \mathbf{0}&  I
          \end{array} \right)
          , \end{equation}
where $I$ is the $3\times3$ identity matrix.\label{symmexplicit}

 Note that since space is isotropic and the
configuration of our systems has one unique divergent direction
$\mathbf{b}$, the vector $\mathbf{e}$  is collinear to
$\mathbf{b}$. Thus
\begin{equation}\label{e}
\mathbf{e}=e\mathbf{b}\end{equation}
 for some constant $e$.
Since the choice of direction of the space coordinate system in
the frame is free, the constant $e$ depends only on $|\mathbf{b}|$
and not on $\mathbf{b}$. Finally, from (\ref{phi}) and (\ref{e}),
it follows that this constant has units (length/time)$^{-2}$.

\begin{figure}[h!]
  \centering
  \scalebox{0.35}{\includegraphics{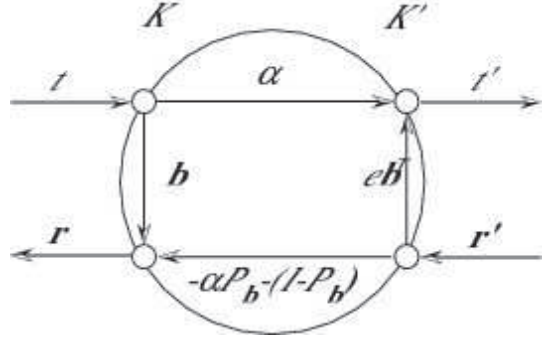}}
  \caption[The explicit form of hybrid connection for space-time transformations.]
  {The hybrid connection for space-time transformations between two inertial
  systems with symmetric frames. The
  circle represents a black box.
   The two input ports are the time $t$ of an event, as measured
 in system $K$, and its space coordinates $\mathbf{r}'$,  as measured
  in system $K'$. The two output ports are
  the time $t'$ of the same event, calculated in system $K'$, and its
  space $\mathbf{r}$ coordinates, calculated in
  $K$. The explicit form of the linear operators representing the functional
  connections between the corresponding ports is shown.
 }\label{hybridend}
\end{figure}

Equation (\ref{sym}) implies
\begin{equation}\label{alpha}
 \alpha=\sqrt{1-e|\mathbf{b}|^2}.
\end{equation}and
\begin{equation}\label{a}
 S_{22}= A=-\alpha P_{\mathbf{b}}-(I-P_{\mathbf{b}}),
\end{equation}
where $P_{\mathbf{b}}$ denotes the orthogonal projection from
space $R^3$ onto the direction of $\mathbf{b}$. Thus, the
space-time transformation between the two inertial frames $K$ and
$K'$ is
\begin{equation}\label{obb}
\left( \begin{array}{c}  t'\\ \mathbf{r}
          \end{array} \right)=
          S_{\mathbf{b}}\left( \begin{array}{c}  t\\ \mathbf{r}'
          \end{array} \right)=
               \left(
         \begin{array}{cc}
              \alpha \quad & e\mathbf{b}^T \\
              \mathbf{b} \quad & -\alpha P_{\mathbf{b}}-(I-P_{\mathbf{b}})
          \end{array} \right)
 \left( \begin{array}{c}  t\\ \mathbf{r}'
          \end{array} \right),
           \end{equation}
with $\alpha$ defined by (\ref{alpha}) (see Figure
\ref{hybridend}).

\begin{figure}[h!]
  \centering
  \scalebox{0.35}{\includegraphics{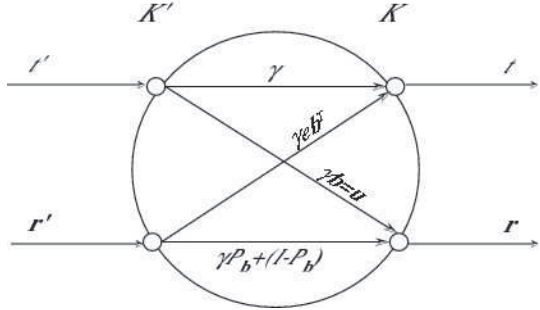}}
  \caption[The explicit form of cascade connection for space-time transformations.]
  {The cascade connection for space-time transformations between two inertial systems
  $K$ and $K'$, moving parallel to $K$ with relative velocity $\mathbf{b}$.
   The two input ports are the time $t'$ of an event and its space
   coordinates $\mathbf{r}'$ , as measured in system $K'$,  and the two output
    ports are the time $t$ of the same event and its
  space $\mathbf{r}$ coordinates, calculated in lab frame $K$.
  The explicit form of the linear operators representing the functional
  connections between the corresponding ports is shown.
 }\label{cascade1end}
 \end{figure}

To compare this result with the usual space-time transformations
in special relativity, we have to recalculate our result for the
cascade connection and reverse the space axes to make them
parallel, as usual. To obtain $\left( \begin{array}{c}  t\\
\mathbf{r}\end{array} \right)$ as a function of $\left(
\begin{array}{c} t'\\\mathbf{r}'\end{array}\right)$
, we use the map $\Psi$ from (\ref{StoE})
and obtain
\begin{equation}\label{cascadespacetimr}
\left( \begin{array}{c}  t\\ \mathbf{r}
          \end{array} \right)=
        \Psi ( S_{\mathbf{b}})\left( \begin{array}{c}  t'\\
        \mathbf{r}'          \end{array} \right)=
             \gamma \left(
         \begin{array}{cc}
              1 & e\mathbf{b}^T \\
              \mathbf{b}&  P_{\mathbf{b}}+\gamma ^{-1}(I-P_{\mathbf{b}})
          \end{array} \right)
 \left( \begin{array}{c}  t'\\ \mathbf{r}'
          \end{array} \right),
\end{equation}
where
\begin{equation}\label{gamma2}
  \gamma =\gamma (\mathbf{b})=1/\sqrt{1-e|\mathbf{b}|^2}.
\end{equation}
This defines an explicit form for the operators of
\index{transformation!space-time} the space-time transformations
using the cascade connection (see Figure \ref{cascade1end}).  If
$e=0$,
 then $\gamma=1$, and the transformations are the\index{Galilean transformation}
\index{transformation!Galilean}  Galilean transformations.

 For the particular case $\mathbf{b}=(v,0,0)$, we get
\begin{equation}\label{coordtrans}
  \begin{array}{cl}
    t &=\gamma(t'+evx') \\
    x & =\gamma(vt'+x') \\
    y & =y' \\
    z & =z'
  \end{array}
\end{equation}
which are the usual Lorentz transformations
\index{transformation!Lorentz}  provided $e=1/c^2$, which will be
shown in the next section.

\section{ Identification of invariants}

In this section we will show that if the transformations are not
Galilean, the principle of relativity \textit{alone} implies that
an interval is conserved and that all velocities are limited by
some universal velocity. To show this, we introduce an appropriate
metric on the space-time under which the symmetry $S_{\mathbf{b}}$
becomes an isometry and a self-adjoint operator. It is known that
$S_{\mathbf{b}}$ is self-adjoint with respect to some inner
product if and only if the eigenvectors of this operator which
correspond to different eigenvalues are orthogonal to each other.

Any symmetry is a reflection with respect to the set of fixed
points. Direct verification shows that the events fixed by the
transformation $S_{\mathbf{b}}$ lie on a straight world-line
through the origin of both frames at time $t=0$, moving with
velocity $\mathbf{w}_1$ (the 1 eigenvector of $S_{\mathbf{b}}$)
defined by
\begin{equation}\label{f2}
  \mathbf{w}_1:=\frac{\mathbf{b}}{
  \alpha +1},
\end{equation}
 where $\alpha$ is defined by (\ref{alpha}). The velocity $\mathbf{w}_1$ is called the \textit{symmetric velocity}
 \index{symmetric velocity}\index{velocity!symmetric} between the
systems $K$ and $K'$. The symmetric velocity has the following
physical interpretation. Place two objects of equal mass (test
masses) at the origins of each inertial system. The center of mass
of the two objects will be called the \index{inertial
system!center} \textit{center of the two inertial
systems}\index{center of two inertial systems}. The symmetric
velocity is the velocity of each system with respect to the center
of the systems.

Similarly, one finds the -1 eigenvectors of $S_{\mathbf{b}}$,
which is defined by
\begin{equation}\label{f3}
  \mathbf{w}_{-1}:=\frac{\mathbf{b}}{
  \alpha -1}.
\end{equation}

The new inner product is obtained by leaving the inner product of
the space components unchanged and introducing an appropriate
weight $\mu$ for the time component. The orthogonality of the
eigenvectors means that
\begin{equation}\label{orth}
 \langle\left( \begin{array}{c}  \mu t\\ \mathbf{w}_1 t
          \end{array} \right)|
          \left( \begin{array}{c} \mu t\\ \mathbf{w}_{-1} t
          \end{array} \right)\rangle=
t ^2(\mu ^2 +\langle\mathbf{w}_1|\mathbf{w}_{-1}\rangle)=0.
\end{equation}
By use of (\ref{f2}), (\ref{f3}) and (\ref{alpha}), this becomes
\begin{equation}\label{mu00}
  \mu ^2+\frac{|\mathbf{b}|^2}{(\alpha+1)(\alpha-1)}=
  \mu ^2+\frac{|\mathbf{b}|^2}{\alpha ^2-1}=\mu
  ^2-\frac{1}{e}=0.
\end{equation}
The orthogonality of the 1 and -1 eigenvectors of $S_{\mathbf{b}}$
is achieved, if $e>0$, by setting
\begin{equation}\label{mu}
  \mu =\frac{1}{\sqrt{e}}.
\end{equation}

In this case, $S_{\mathbf{b}}$ becomes an isometry with respect to
the inner product with weight $\mu$, implying that
\begin{equation}\label{mu1}
(\mu t)^2 + |\mathbf{r}'|^2=(\mu t')^2 + |\mathbf{r}|^2,
\end{equation}
or, equivalently,
\begin{equation}\label{intconserv}
  (\mu t')^2 - |\mathbf{r}'|^2=(\mu t)^2 - |\mathbf{r}|^2.
\end{equation}

 The previous equation implies that our space-time transformation
from $K$ to $K'$ conserves the \textit{relativistic interval}
\index{interval!space-time}\index{Lorentz
invariant!interval!space-time}
\begin{equation}\label{interval1}
ds^2=(\mu dt)^2 - |d\mathbf{r}|^2,
\end{equation}
with $\mu$ defined by (\ref{mu}) and determined be the process of
synchronization of the clocks.

 In particular, the
transformation $S_{\mathbf{b}}$ maps zero interval world-lines to
zero interval \index{zero interval}
\index{interval!zero}world-lines. Since zero interval world-lines
correspond to uniform motion with unique speed $\mu$, for any
relativistic space-time transformation between two inertial
systems with $e>0$, there is a speed $\mu$ defined by (\ref{mu})
which is conserved. Obviously, the cone $ds^2>0$, corresponding to
the positive\index{Lorentz!cone} \textit{Lorentz cone}, is also
preserved under this transformation.

It can be shown that $e$ is independent of the relative velocity
$\mathbf{b}$ between the frames $K$ and $K'$.

 Several
experiments at end of 19th century  showed that the speed of light
is the same in all inertial systems. Thus
\begin{equation}\label{muandc}
 \mu =c\;\;\mbox{ and }\;\; e=\frac{1}{c^2},
\end{equation}
where $c$ is the \textit{speed of light}\index{speed of light} in
a vacuum. Based on this, we can rewrite (\ref{alpha}) and
(\ref{gamma2}) as
\begin{equation}\label{gammaalpha}
  \alpha
  (\mathbf{b})=\sqrt{1-\frac{|\mathbf{b}|^2}{c^2}},\;\gamma(\mathbf{b})=\frac{1}{\sqrt{1-\frac{|\mathbf{b}|^2}{c^2}}}
\end{equation}
 and  the space-time
transformations (\ref{cascadespacetimr}) and (\ref{coordtrans})
between two inertial systems are the Lorentz transformations.

 If  $e=0$, the
above space-time transformations become the Galilean
transformations and in this case no velocity is preserved. It can
be shown that the case $e<0$ leads to physically absurd results,
leaving only two possibilities for relativistic space-time
transformations: the Galilean and Lorentz transformations.

Since in the Minkowski metric $\mathbf{w}_1$ is orthogonal to
$\mathbf{w}_{-1}$, the matrix $S_\mathbf{b}$ is self-adjoint.
Therefore, the \textit{adjoint of the relative velocity}
$\mathbf{b}$, as a linear operator from time to space, is the
operator of \textit{non-simultaneity} of a system $K'$ moving with
relative velocity $\mathbf{b}$ with respect to the lab frame.

\section{Velocity addition and symmetry of the velocity ball}

Relativistic velocity addition can be derived from the Lorentz
space-time transformation (\ref{cascadespacetimr}) between two
inertial systems as follows. Consider two inertial systems  $K$
and $K'$, moving with relative velocity (boost) $\mathbf{b}$ and
 a motion with
 uniform velocity $\mathbf{v}$ in system $K'$. The world line of
 this motion is $\left(
\begin{array}{c} t'\\ \mathbf{v}t'\end{array} \right)$ in  $K'$.
From (\ref{cascadespacetimr})
and by use of (\ref{muandc}) this world line in system $K$ is
 \[ \gamma\left(
         \begin{array}{c}
              t'+  \frac{\mathbf{b}^T\mathbf{v}t'}{c^2} \\
              \mathbf{b} t'+ t'P_{\mathbf{b}}\mathbf{v}+\alpha t'(I- P_{\mathbf{b}})\mathbf{v}
          \end{array} \right)
 \]
 or
 \begin{equation}\gamma t' \left(
         \begin{array}{c}
             1 +  \frac{\langle\mathbf{b}|\mathbf{v}\rangle}{c^2} \\
              \mathbf{b} +  \mathbf{v}_{\|}+\alpha \mathbf{v}_\bot
          \end{array} \right),\end{equation}
          where $\mathbf{v}_{\|}=P_{\mathbf{b}}\mathbf{v}$ denotes
          the component of $\mathbf{v}$ parallel to $\mathbf{b}$ and
          $\mathbf{v}_\bot=(I-P_{\mathbf{b}})\mathbf{v}$ denotes
          the component of $\mathbf{v}$ perpendicular  to $\mathbf{b}.$
The world line in system $K$ is a straight line corresponding to
the velocity, called the \textit{relativistic velocity sum}
$\mathbf{b} \oplus \mathbf{v}.$ This velocity is obtained by
dividing the space by the time on the line and is
\begin{equation}\label{veladd}
  \mathbf{b} \oplus \mathbf{v}=\frac{ \mathbf{b} +  \mathbf{v}_{\|}+\alpha \mathbf{v}_\bot}{1 +
   \frac{\langle\mathbf{b}|\mathbf{v}\rangle}{c^2}},
\end{equation}
with $\alpha =\alpha (\mathbf{b})=\sqrt{1-|\mathbf{b}|^2/c^2}$.
This is the well-known Einstein velocity addition formula.

In case  $\mathbf{b}$ and  $ \mathbf{v}$ are parallel,
 this formula becomes:
\begin{equation}\label{veladdpar}
  \mathbf{b} \oplus \mathbf{v}=\frac{\mathbf{b} +
  \mathbf{v}}{1 +
   \frac{{b}{v}}{c^2}},
\end{equation}
and in case  $\mathbf{v}$ is  perpendicular to $ \mathbf{b}$ the
formula becomes:
\begin{equation}\label{veladdorth}
  \mathbf{b} \oplus \mathbf{v}=\mathbf{b} +\alpha(\mathbf{b})
  \mathbf{v}.
\end{equation}
Note that the velocity addition is commutative only for parallel
velocities.

We denote by $D_v$ the set of all relativistically admissible
velocities in an inertial frame $K$. This set is defined by
\begin{equation}\label{velballdef}
 D_v=\{\mathbf{v}: \; \mathbf{b}\in R^3, \;|\mathbf{b}|<c\}.
\end{equation}
The Lorentz transformation (\ref{cascadespacetimr}) acts on the
velocity ball $D_v$ as
\begin{equation}\label{boostvelball}
 \varphi _{\mathbf{b}}(\mathbf{v})=\mathbf{b} \oplus \mathbf{v}=
 \frac{ \mathbf{b} +  \mathbf{v}_{\|}+\alpha \mathbf{v}_\bot}{1 +
   \frac{\langle\mathbf{b}|\mathbf{v}\rangle}{c^2}}.
\end{equation}
 It can be shown \cite{F04}
that the map $\varphi _{\mathbf{b}}$ is a projective (preserving
line segments) map of $D_v$. We denote by $Aut _p(D_v)$ the group
of all projective automorphisms of the domain $D_v$. The map
$\varphi _{\mathbf{b}}$ belongs to $Aut _p(D_v).$ It transforms
any relativistically admissible  velocity $\mathbf{v}\in D_v$ of
the system $K'$, which is moving parallel to $K$ with relative
velocity $\mathbf{b}$, to a corresponding unique velocity
$\varphi_{\mathbf{b}}(\mathbf{v})\in D_v$ in $K$. The existence of
such a map shows that the ball $D_v$ is homogeneous in the sense
that any two points of this ball can be exchanged by an element of
$Aut _p(D_v).$ Moreover, the ball $D_v$ is a bounded symmetric
domain, meaning that for any point in $D_v$ there exist a symmetry
belonging to $Aut _p(D_v)$ which fixes only this point.

 By a simple argument one can show that the group $Aut
_p(D_v)$ of all projective automorphisms is
\begin{equation}\label{AutD0}
 Aut _p(D_v)=\{\varphi_{\mathbf{b},U}=\varphi_{\mathbf{b}}U :
  \mathbf{b} \in D_v, \; U\in O(3)\},
\end{equation}
where $O(3)$ denotes the orthogonal matrix of size $3\times 3$.
This group represents the velocity transformation between two
arbitrary (even non-parallel) inertial systems and provides a
representation of the Lorentz group.

 Note that the Lorentz group representation defined by space-time
 transformations (\ref{cascadespacetimr}) between two inertial systems
is valid only if the systems move in parallel and at time $t=0$
the origins of the two systems coincide, while the velocity
transformation (\ref{AutD0}) between two inertial systems holds
for arbitrary systems without any limitation.

\section{Kinematics of relativistically accelerated systems}

\begin{figure}[h!]\begin{center}
  \scalebox{0.6}{\includegraphics{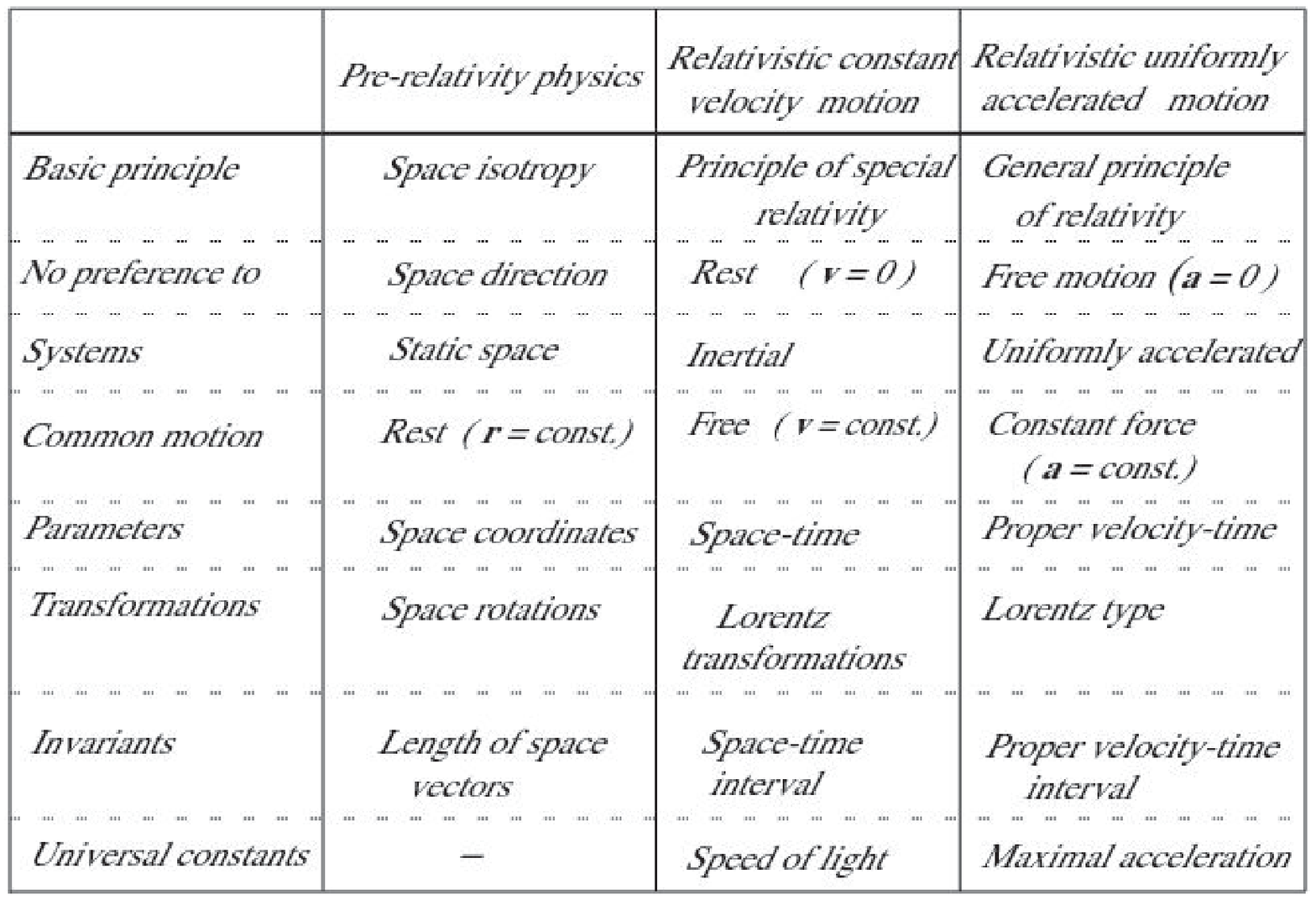}}
 \end{center} \caption{ Comparisons between inertial and
 uniformly accelerated systems.}\label{table}
\end{figure}

The ideas used for inertial systems can also be applied with some
modifications to uniformly accelerated systems. To understand why
we are justified in applying our method used for the inertial
systems to \textit{accelerated} systems and what modifications are
needed, consider the table in Figure \ref{table}, which clarifies
our line of reasoning.
It highlights three areas of physics:
\begin{itemize}
  \item Pre-relativity physics, which deals with static
  space
  \item Relativistic physics, which deals with arbitrary
  inertial systems
  \item Relativistic physics, which deals with systems that
   are accelerated with respect to inertial systems.
\end{itemize}

Each of these three  areas mentioned has its basic principle which
states that the laws of physics are independent of a particular
\textit{choice}. In pre-relativity physics, the laws are
independent of the choice of a preferred space direction. In
special relativity, the laws are independent of the choice of a
preferred inertial system. \index{system!prefered} In general
relativity, they are independent of the choice of an arbitrary
system.

Pre-relativity physics abandons the notion of a preferred space
direction but maintains a preference for rest $(\mathbf{v}=0)$.
Special relativity, describing uniform motion, abandons the
preference for rest but maintains a preference for constant
velocity $(\mathbf{a}=0)$. Here the preferred type of motion is
free motion - a motion which is free in one inertial system is
free in \textit{every} inertial system. For accelerated motion,
the Principle of General relativity abandons the preference for
constant velocity. Even though there is absolutely no preference
for any particular kind of accelerated system, we will give
preference to \textit{uniformly accelerated systems}, meaning
systems that are uniformly accelerated with respect to an inertial
system.  The preferred type of motion will thus be motion under a
constant force.

For relativistic constant velocity motion, we use the space-time
description of events because it is the simplest one which leads
to \textit{linear} transformations between inertial systems.
However, it is impossible to obtain linear transformations between
uniformly accelerated systems using the space-time description.
Consequently, for accelerated comoving systems, we will use a new
description, called the \textit{proper velocity-time
description}.\index{description!proper velocity-time}
 This will enable us to obtain\index{transformation!linear}
\textit{linear} transformations between comoving uniformly
accelerated systems.

Each of these three areas of physics has its own invariants. For
constant velocity motion, our method \cite{F04} produced
transformations which preserve a space-time interval and a maximal
speed $c$- the speed of light. Similarly, for uniformly
accelerated motion, our method will produce transformations which
preserve a proper velocity-time \index{interval!proper
velocity-time}interval and a \textit{maximal
acceleration}.\index{maximal acceleration}

\subsection{Proper acceleration}

From (\ref{propervelocity}), (\ref{teta}) and (\ref{gammaalpha})
the \textit{proper velocity} \index{proper velocity} of an object
$\mathbf{u}$ is
\begin{equation}\label{tm}
\mathbf{u}=\frac{d\mathbf{r}}{d\tau}=\gamma
(\mathbf{v})\mathbf{v}=
\frac{\mathbf{v}}{\sqrt{1-{|\mathbf{v}|^2}/{c^2}}},
\end{equation}
  where
$d\tau=\sqrt{1-{|\mathbf{v}|^2}/{c^2}}dt$ is the proper time
interval, \textit{i.e.}, the time interval in the frame moving
with the object. For brevity, we will call proper velocity
\textit{p-velocity}. Note that a p-velocity is expressed as a
vector of $R^3$. Conversely, any vector in $R^3$, with no
limitation on its magnitude, represents a relativistically
admissible p-velocity.

The principle of equivalence \index{principle!equivalence}states
that ``the laws of physics have the same form in a uniformly
accelerated system as they do in an unaccelerated inertial system
in a uniform gravitational field." But what is the meaning of
uniform acceleration in this principle? Consider an inertial
system in a constant gravitational field $G$. If we position an
object freely in this system, then, from the relativistic dynamic
equation, which we write as
\begin{equation}\label{propevolution}
 m_0 \frac{d\mathbf{u}}{dt}=G,
\end{equation}
we obtain that $\frac{d\mathbf{u}}{dt}$ remains constant. We
define \textit{ proper acceleration} $\mathbf{g}$ to be the
derivative of p-velocity with respect to time $t$, \textit{i.e.},
\begin{equation}\label{acceldef}
 \mathbf{g}=
\frac{d\mathbf{u}}{dt}=\frac{d^2\mathbf{r}}{dt d\tau}.
\end{equation}
This definition coincides with the one given in \cite{rindler}
p.71. By this definition, a free object in an inertial system with
a constant gravitational field has constant proper acceleration.
By the Equivalence Principle a uniformly accelerated system has to
be one in which the origin moves with constant proper
acceleration.
 The magnitude of  the proper acceleration is larger then the usual acceleration
$\frac{d^2\mathbf{r}}{dt^2}$, but for small velocities, the
magnitudes are almost the same.

By \textit{uniformly accelerated system } in this paper we will
mean systems that are moving with \textit{constant proper
acceleration} with respect to a given inertial system. As
mentioned earlier, we will describe first the transformation
between an inertial system and uniformly accelerated system (in
sense of constant \textit{proper} acceleration).

\subsection{Proper velocity - time description of
events}\label{sec.1.6.2}

An important step in our derivation of the Lorentz space-time
transformations between two inertial frames was to show that such
transformations are linear. For uniformly accelerated systems, the
space-time transformation is not linear. But, there is another
description of events, called the \textit{proper velocity - time}
description, in which the transformation of events between two
uniformly accelerated systems is linear.

 In the p-velocity-time description, an event is
described by the time at which the event occurred and the
p-velocity $\mathbf{u}\in R^3$ of the event. Thus, an event in the
p-velocity-time description is described by a vector
 $\left(  \begin{array}{c} t\\ \mathbf{u}\end{array} \right)$ in $R^4$.
 The evolution of an object is described by the
p-velocity $\mathbf{u}(t)$ of the object at time $t$, replacing
the world-line of special relativity. To obtain the position of
the object at time $t$, we have to know the initial position of
the object and then integrate its velocity, expressed uniquely by
the p-velocity, with respect to time.

For example, for a free-falling object which at time $t=0$ was at
rest and positioned at the space frame origin, the p-velocity of
this object at time $t$ is $\mathbf{u}(t)=\mathbf{g}t$, where
$\mathbf{g}$ is the proper acceleration generated by the
gravitational field.

By the principle of equivalence and (\ref{propevolution}), the
motion of an object under the influence of a constant force in a
uniformly accelerated system is equivalent to its motion in a
constant gravitational field. This motion is described by a
straight line in the p-velocity-time
 continuum. Conversely, if the motion of an object in the p-velocity-time
 continuum of a uniformly accelerated system is described by a straight
 line, then the object is under the influence of a constant force.

Two systems moving parallel to each other are called
\textit{comoving} if at some initial time $t_0$ their relative
velocity is zero. Consider now two comoving systems  $K_g$ and
$K_0$, uniformly accelerated  with respect to an inertial system
$K$ with a constant acceleration $\mathbf{g}$ between them. We
assume that at time $t=0$ the  relative velocity of $K_0$ with
respect to $K_g$ is zero.

Denote by $T$ the transformation mapping
 the time and p-velocity $\left(\begin{array}{c} t\\
 \mathbf{u}\end{array} \right)$ of an event, measured in $K_g$, to the
 time and p-velocity of the same event $\left(\begin{array}{c} t'\\
 \mathbf{u}'\end{array} \right)$, measured in $K_0$.
As we have stated, motion under a constant force in one uniformly
accelerated system with respect to an inertial system $K$ is
equivalent to motion under a different, but constant, force in
$K$. Thus, motion under a constant force in one accelerated system
with respect to an inertial system $K$ will be of the same type in
any other system uniformly accelerated with respect to $K$. Since
such motion is described by a straight line in p-velocity time
continuum, this implies that the map $T$ preserves straight lines.
We also assumed that relative velocity of $K_0$ with respect to
$K_g$ is zero at time $t=0$. Thus, by a known theorem in
mathematics, maps preserving straight lines and mapping the origin
to the origin are linear. Since $T$ satisfies these conditions,
\index{transformation!linear} the transformation $T$ is a
\textit{linear map}.

\subsection{Identification of symmetry}\label{sec.1.6.3}

 To define the symmetry operator  between two
uniformly accelerated systems, we will use an extension of the
principle of relativity, \index{relativity!principle!general}
which we will call the \textit{General Principle of
Relativity}.\index{principle!relativity!general} This principle,
as it was formulated by M. Born (see \cite{B65}, p. 312), states
that the ``laws of physics involve only relative positions and
motions of bodies. From this it follows that no system of
reference may be favored \textit{a priori} as the inertial systems
were favored in special relativity."

 The principle of relativity
from special relativity states that there is no preferred
\textit{inertial} system, and, therefore, the notion of rest (zero
velocity) is a relative notion. The motions which were common for
all inertial systems are the free (constant velocity) motions.
From the general principle of relativity, it follows that there is
no preference for inertial (zero acceleration) systems. Hence,
when considering accelerated systems, we no longer give preference
to free motion (zero force) over constant force motion. This makes
all uniformly accelerated systems equivalent. The motion which is
common for all accelerated systems is motion under a constant
force.

 From the general principle of relativity, it is logical
to assume that \textit{the transformations between the
descriptions of an event in two uniformly accelerated systems
depend only on the relative motion between these systems}. Thus,
if we choose reference frames in the two uniformly accelerated
systems  $K_g$ and $K_0$ in a way that the description of relative
motion of $K_0$ with respect to  $K_g$ coincides with the
description of relative motion of  $K_g$ with respect to $K_0$,
then the transformation $T$, defined above, will coincide with the
transformation $\widetilde{T}$ from  $K_0$ to  $K_g$. This implies
that $T$ is a symmetry.

 In order for this transformation to be a
  symmetry,  we have to choose the space axes
 in such a way that the description of the relative position of system
 two with respect to system one coincides with the description of
 the relative position of system one with respect to system two.
 This can be done in the same way as was done for inertial
 systems in Section \ref{synchron}. By reversing the p-velocity
 axes in $K_0$ we can make the relative acceleration of
 $K_g$ with respect to $K_0$ also be $\mathbf{g}$.

 In order to describe the precise meaning of ``the system $K_0$
moves with uniform acceleration $\mathbf{g}$ with respect to
$K_g$," we consider an event connected to an object which is at
rest at $O'$-the origin of $K_0$. The p-velocity of this object in
$K_0$ is $\mathbf{u}'=0$. The acceleration of system $K_0$ with
respect to $K_g$ expresses the p-velocity $\mathbf{u}$ of this
object after time $t$ in $K_g$.  Thus, p-velocity in system $K_0$
and time in $K_g$ served as inputs and p-velocity in $K_g$ as an
output. This corresponds to the hybrid connection  \index{hybrid
connection} discussed in Section \ref{hybridsec}.

 The p-velocity-time transformation $S_\mathbf{g}$
between these frames can be considered as a ``two-port linear
black box" transformation with two inputs and two outputs. In each
system, the time and the p-velocity of an event are the two ports.
We choose the inputs to be the scalar $t$, the time of the event
in $K_g$, and the three-dimensional vector $\mathbf{u}'$
describing the p-velocity of the event in $K_0$. Then the outputs
are the scalar $t'$, the time of the event in $K_0$, and the
three-dimensional vector $\mathbf{u}$ describing the p-velocity of
the event in $K_g$. We denote by $S_\mathbf{g}$ mapping
 $\left(\begin{array}{c} t\\
 \mathbf{u}'   \end{array} \right)$ to $\left(\begin{array}{c} t'\\
  \mathbf{u}  \end{array} \right),$ which is uniquely defined by $T$. The symmetry of the
p-velocity-time transformation $T$ implies (see \cite{F04}) that
also the transformation $S_\mathbf{g}$ is a symmetry and the
linearity of $T$ implies the linearity of the transformation
$S_\mathbf{g}$.

\subsection{Identification of the transformation}\label{sec.1.6.6}

 The map $S_\mathbf{g}$ can be represented by a $4\times 4$ matrix.
The four block components of the transformation $S_\mathbf{g}$,
defined by
\begin{equation}\label{comp1a}
 \left( \begin{array}{c}  t'\\ \mathbf{u}
          \end{array} \right)=S_\mathbf{g}\left( \begin{array}{c}  t\\ \mathbf{u}'
          \end{array} \right)= \left(
         \begin{array}{rr}
              S_{11} & S_{12} \\
              S_{21}& S_{22}
          \end{array} \right)
          \left( \begin{array}{r}  t\\ \mathbf{u}'
          \end{array} \right),
\end{equation}
 will be
denoted by $S_{ij}$, for $i,j\in \{1,2\}$, as in Figure
\ref{Fhybrid}.

Obviously, $S_{21}=\mathbf{g}$. The map $S_{11}$ describes the
transformation of the time $t$ in $K_g$ of an event with
p-velocity $\mathbf{u}'=0$ (at rest in $K_0$) to its time $t'$ in
$K_0$, and it is given by
\begin{equation}\label{tetap}
    t'=S_{11}( t)=\beta  t,
\end{equation}
for some constant $\beta$. The constant $\beta$ expresses the
slowdown  of the clocks at rest in $K_0$ due to its acceleration
relative to $K_g$. The value of $\beta$ is related to the
well-known  \textit{Clock Hypothesis}.\index{clock hypothesis}
Recall that the Clock Hypothesis states\index{accelerated clock
rate} that the ``rate of an accelerated clock is identical to that
of instantaneously comoving inertial clock." Now the p-velocity of
$K_0$ with respect to $K_g$ was zero at time $t=0$. Therefore, at
time $t=0$, both $K_0$ and $K_g$ have the same comoving inertial
system. Thus, the clock hypothesis would imply $\beta =1$.

To define the maps $S_{12}$ and $S_{22}$, we will consider an
event occurring  at time $t=0$ in $K_g$ with p-velocity $
\mathbf{u}'$ measured in $K_0$. Then $S_{12} (\mathbf{u}')$ will
be the time of this event in $K_0$, and $S_{22} (\mathbf{u}')$
will be the p-velocity of this event in $K_g$. Since $S_{12}$ is a
linear map from $R^3$ to $R$, we have
\begin{equation}\label{phip}
  S_{12}(\mathbf{u}')=<\mathbf{h}|\mathbf{u}'>=
  {\mathbf{h}^T}\cdot\mathbf{u}',
\end{equation}
for some vector $\mathbf{h}\in R^3$. $S_{12}(\mathbf{u}')$
measures the non-synchronization in $K_0$ at $t=0$ of two clocks,
one at rest and one moving with constant p-velocity $\mathbf{u}'$
in $K_0$, where both clocks were synchronized at $t=0$ in system
$K_g$.

 Note that since space is isotropic and the
configuration of our systems has one unique divergent direction
$\mathbf{g}$, the vector $\mathbf{h}$  is collinear to
$\mathbf{g}$. Thus
\begin{equation}\label{eep}
\mathbf{h}=\kappa\mathbf{g},
\end{equation}
 for some constant $\kappa$.
Since the choice of direction of the space coordinate system  in
the frame is free, the constant $\kappa$ depends only on
$|\mathbf{g}|$ and not on $\mathbf{g}$. From (\ref{phip}) and
(\ref{eep}), it follows that this constant has units
$(length/time^2)^{-2}$.

The map $S_{22}$ describes the p-velocity difference in  $K_g$ of
simultaneous events in $K_g$ with respect to their p-velocity
difference in $K_0$, and it is given by
\begin{equation}\label{gammap}
  S_{22} (\mathbf{u}')=A\mathbf{u}'
\end{equation}
for some $3\times 3$ matrix $A$.

 Our black box transformation can now be described
by a $4 \times 4$ matrix $S_\mathbf{g}$ with block matrix entries
from (\ref{tetap}), (\ref{phip}), (\ref{eep}) and (\ref{gammap}),
as
\begin{equation}\label{bbtp}
S_\mathbf{g}=
          \left(
         \begin{array}{rr}
              \beta & \kappa\mathbf{g}^T \\
              \mathbf{g}&  A
          \end{array} \right)
          . \end{equation}
 Since $S_\mathbf{g}$ is a symmetry operator, it follows that
 $S_\mathbf{g}^2=I$-the identity. Hence,
 \begin{equation}\label{symp}
\left(         \begin{array}{cc}
              \beta & \kappa\mathbf{g}^T \\
              \mathbf{g}&  A
          \end{array} \right)
          \left( \begin{array}{rr}
           \beta & \kappa\mathbf{g}^T \\
              \mathbf{g}&  A
          \end{array} \right)=\left( \begin{array}{rr}
           1 & \mathbf{0}^T \\
              \mathbf{0}&  I
          \end{array} \right)
          , \end{equation}
where $I$ is the $3\times 3$ identity matrix.

 From the
multiplication of the first row by the first column we have $
\beta^2 +\kappa|\mathbf{g}|^2=1$ and, thus
\begin{equation}\label{1'p}
 \beta=\sqrt{1-\kappa|\mathbf{g}|^2}.
\end{equation}
 From the
multiplication of the last 3 rows row by the last 3 columns we
have $ \kappa\mathbf{g}\mathbf{g}^T+A^2=I.$ Using that
$\mathbf{g}\mathbf{g}^T=|\mathbf{g}|^2P_\mathbf{g},$ where
$P_\mathbf{g}$ denotes the orthogonal projection on the direction
of $\mathbf{g},$ we get $A^2=I-\kappa|\mathbf{g}|^2P_\mathbf{g}=
\beta ^2P_{\mathbf{g}}-(I-P_{\mathbf{g}}).$ Thus,
\begin{equation}\label{ap}
  A=-\beta P_{\mathbf{g}}-(I-P_{\mathbf{g}}).
\end{equation}
The negative sign is chosen because of the space reversal of axes.
Thus, the p-velocity-time transformation between the two frames
$K_g$ and $K_0$ is
\begin{equation}\label{obbp}
\left( \begin{array}{c}  t'\\ \mathbf{u}
          \end{array} \right)=
          S_\mathbf{g}\left( \begin{array}{c}  t\\ \mathbf{u}'
          \end{array} \right)= \left(
         \begin{array}{cc}
              \beta & \kappa\mathbf{g}^T \\
              \mathbf{g}& -\beta P_{\mathbf{g}}-(I-P_{\mathbf{g}})
          \end{array} \right)
 \left( \begin{array}{c}  t\\ \mathbf{u}'
          \end{array} \right),
           \end{equation}
with $\beta$ defined by (\ref{1'p}).

Next, to define an explicit form for the operators of
\index{transformation!p-velocity-time} the p-velocity-time
transformations between two comoving uniformly accelerated systems
using the cascade connection, we use the map $\Psi$ from
(\ref{StoE}) and revers the space axes to make them parallel, as
usual. We obtain
\begin{equation}\label{cascadespacetimr1}
\left( \begin{array}{c}  t\\ \mathbf{u}
          \end{array} \right)=
             \beta^{-1} \left(
         \begin{array}{cc}
              1 & \kappa\mathbf{g}^T \\
              \mathbf{g}&  P_{\mathbf{g}}+\beta(I-P_{\mathbf{g}})
          \end{array} \right)
 \left( \begin{array}{c}  t'\\ \mathbf{u}'
          \end{array} \right).
\end{equation}

To compare these transformations with the well known Lorentz
transformations, we choose the $x$-axis of $K_g$ in the direction
of $\mathbf{g}$. Denote $\mathbf{g}=(g,0,0)$,
$\mathbf{u}=(u_x,u_y,u_z)$ and $\mathbf{u}'=(u'_x,u'_y,u'_z)$. We
get
\begin{equation}\label{coordtransp}
  \begin{array}{cl}
    t &=\beta^{-1}(t'+\kappa gu'_x) \\
    u_x & =\beta^{-1}(gt'+u'_x) \\
    u_y & =u'_y \\
    u_z & =u'_z,
  \end{array}
\end{equation}
which is a\textit{ Lorentz-type transformation } of the
p-velocity.\index{Lorentz!transformation! accelerated systems}

If $\kappa=0$, which corresponds to the assumption of the Clock
Hypothesis, then from (\ref{1'p}) we have $\beta =1$ and the
p-velocity-time transformations (\ref{coordtransp}) become
Galilean.

 From now on we will consider only the case $\kappa\neq
0$.

\section{ Conservation of p-velocity time interval and maximal acceleration}\label{sec.1.6.7}

\begin{figure}[h!]
   \begin{picture}(200,120)(-80,0)
   \put (45,10){\vector(1,0){45}}
   \put (90,10){\vector(1,0){45}}
   \put (135,10){\vector(1,0){20}}
   \put (45,00){\vector(0,1){120}}
   \put (45,10){\vector(-1,0){30}}
   \put (45,10){\line(-1,3){35}}
   \put (45,10){\line(1,1){95}}
   \put (45,10){\line(1,2){47}}
   \put (43,100){\line(1,0){4}}
   \put (15,10){\dashbox(120,90){ }}
   \put (90,10){\dashbox(00,90){ }}
   \put (47,123){$t$}
   \put (47,3){$0$}
   \put (92,3){$\mathbf{w}_1$}
   \put (13,3){$\mathbf{w}_{-1}$}
   \put (37,103){$1$}
   \put (89,48){$\mathbf{u}=\mathbf{g}t$}
   \put (137,3){$\mathbf{g}$}
   \put (-13,45){$\mathbf{u}=\mathbf{w}_{-1}t$}
   \put (45,85){$\mathbf{u}=\mathbf{w}_1t$}
     \end{picture}
  \caption{Eigenspaces of the symmetry.}\label{eigenspacep}
\end{figure}
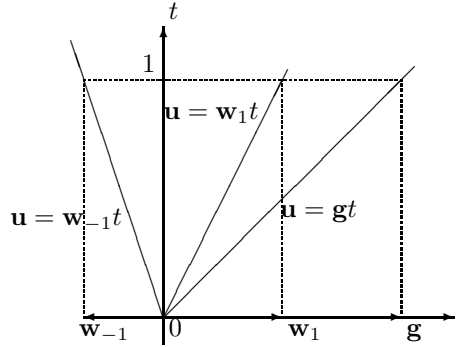

As mentioned above, the p-velocity-time transformation between the
systems $K_g$ and $K_0$ is a symmetry transformation. Such a
symmetry is a reflection with respect to the set of points fixed
by the symmetry. The  events fixed by the transformation
$S_\mathbf{g}$ are on a straight line through the origin of the
p-velocity-time continuum, corresponding to the motion of an
object with constant acceleration $\mathbf{w}_1$ (see Figure
\ref{eigenspacep}) in both frames, where $\mathbf{w}_1$ is
\begin{equation}\label{f2p}
  \frac{\mathbf{g}}{1
  +\beta}=\frac{\mathbf{u}}{t}:=\mathbf{w}_1.
\end{equation}

The events, which are the -1 eigenvectors of $S_\mathbf{g}$ in the
plane generated by $\mathbf{g}$ and the $t$-axis, are on a
straight line through the origin of the p-velocity-time continuum,
corresponding to the motion of an object with constant
acceleration $\mathbf{w}_{-1}$, defined as
\begin{equation}\label{f3p}
  \frac{\mathbf{u}'}{t}=\frac{\mathbf{g}}{
  \beta -1}=\frac{\mathbf{u}}{t}:=\mathbf{w}_{-1}.
\end{equation}

The symmetry $S_\mathbf{g}$ becomes an isometry if we introduce an
appropriate inner product. Under this inner product, the 1 and -1
eigenvectors of $S_\mathbf{g}$ will be orthogonal. The new inner
product is obtained by leaving the inner product of the p-velocity
components unchanged and introducing an appropriate weight $\nu$
for the time component. The orthogonality of the eigenvectors
means that
\begin{equation}\label{orth}
 <\left( \begin{array}{c}  \nu t\\ \mathbf{w}_1 t
          \end{array} \right)|
          \left( \begin{array}{c} \nu t\\ \mathbf{w}_{-1} t
          \end{array} \right)>
=t ^2(\nu ^2 +<\mathbf{w}_1|\mathbf{w}_{-1}>)=0.
\end{equation}
By use of (\ref{f2p}), (\ref{f3p}) and (\ref{1'p}), this becomes
\begin{equation}\label{mu00p}
  \nu ^2+\frac{|\mathbf{g}|^2}{(1+\beta)(\beta-1)}=
  \nu ^2-\frac{1}{\kappa}=0.
\end{equation}

If $\kappa>0$, this implies that
\begin{equation}\label{mup}
  \nu =\frac{1}{\sqrt{\kappa}}.
\end{equation}
The value $\nu$ has units of acceleration. From the fact that
$S_\mathbf{g}$ is an isometry with respect to the inner product
with weight $\nu$, we have
\begin{equation}\label{mu1p}
(\nu t)^2 + |\mathbf{u}'|^2=(\nu t')^2 + |\mathbf{u}|^2,
\end{equation}
or, equivalently,
\begin{equation}\label{intconservp}
  (\nu t')^2 - |\mathbf{u}'|^2=(\nu t)^2 - |\mathbf{u}|^2.
\end{equation}
The previous equation implies that our p-velocity-time
transformation from $K_g$ to $K_0$ conserves the interval
\begin{equation}\label{interval2}
d\widetilde{s}^2=(\nu dt)^2 - |d\mathbf{u}|^2,
\end{equation}
with $\nu$ defined by (\ref{mup})\index{interval!proper
velocity-time} \index{Lorentz
invariant!interval!p-velocity-time}and thus is a Lorentz-type
transformations.

 Note that the zero-interval world-lines are
transformed by these transformations to zero-interval lines. The
zero-interval world-lines correspond to motion with uniform
acceleration $\nu$ depending only on the magnitude $g$ of the
relative accelerations between the systems, which we denote by
$\nu _g$. Thus, for two systems $K_g$ and $K_0$ with $\kappa>0$,
the acceleration $\nu _g$ defined by (\ref{mup}) is conserved.

Considering three accelerated comoving systems: where system two
is moving in parallel with acceleration $\mathbf{g}$ with respect
to system one and system three is moving in parallel with the same
acceleration $\mathbf{g}$ with respect to system two we obtain
that $\nu _{2g}=\nu _g$. By use of an argument similar to the one
in \cite{F04}, section 1.2.2, it can be shown that the conserved
acceleration $\nu _g$ is independent of the relative acceleration
$\mathbf{g}$ between the frames $K_g$ and $K_0$ and we will denote
it by $a_M$. From (\ref{mup})
\begin{equation}\label{maxaccelp}
 a_M =1/\sqrt{\kappa} .
\end{equation}\index{acceleration!maximal}
Since the cone $d\widetilde{s}^2>0$ is preserved under the
p-velocity-time transformation, an acceleration of magnitude less
them $a_M$ in one uniformly accelerated system will be also of
magnitude less them $a_M$ in any other uniformly accelerated
system. Thus, all relativistically admissible accelerations,
\index{ball!relativistically admissible!accelerations} belong to a
ball $D_a$
\begin{equation}\label{accelball}
 D_a=\{ \mathbf{a}\in R^3:\;\;|\mathbf{a}|<a_M \}.
\end{equation}

The case $\kappa<0$ is excluded by an argument, similar to the one
used in \cite{F04}, section 1.3.2.

\section{The ball of relativistically admissible accelerations $D_a$}\label{sec.1.6.8}

If the maximal acceleration exists, the constant $\kappa=1/a_M^2$
and the p-velocity-time transformations (\ref{coordtransp}) become
\begin{equation}\label{coordtranspfinal2}
  \begin{array}{cl}
    t &=\beta^{-1}(t'+ \frac{gu'_x}{a_M^2}) \\
    u_x & =\beta^{-1}(gt'+u'_x) \\
    u_y & =u'_y \\
    u_z & =u'_z,
  \end{array}
\end{equation}
with $\beta=\sqrt{1-\frac{g^2}{a_M^2}}.$ From this one derives a
new acceleration addition formula as follows. Consider an object
moving with acceleration $\mathbf{a}=(a,0,0)$ in the direction of
the of the $x'$ axis in $K_0$ with zero p-velocity at $t'=0$. Its
p-velocity in $K_0$ is $u'_x=at',u'_y=0,u'_z=0$. Thus, the
p-velocity-time description of this object in system $K_g$ is
$t=\beta^{-1}t'(1+ \frac{ga}{a_M^2}),u_x =\beta^{-1}t'(g+a),u_y
=0, u_z=0$. Such motion of this object in $K_g$ represents
addition of the acceleration $\mathbf{g}$ of the system $K_0$ with
respect to $K_g$ to the acceleration $\mathbf{a}$ of the object
with respect to $K_0$. Since,
\[\frac{u_x}{t}=\frac{g+a}{1+ \frac{ga}{a_M^2}}\mbox{ and }\;\frac{u_y}{t}=\frac{u_z}{t}=0,\]
the motion is with constant acceleration. This operation is
denoted by $\mathbf{g}\oplus \mathbf{a}$ and coincides with the
similar formula of Einstein velocity addition (for parallel
velocities) in special relativity.

A general acceleration-addition formula for non-parallel
accelerations is
\begin{equation}\label{acceladdp}
 \mathbf{g}\oplus  \mathbf{a}=
  \frac{ \mathbf{g}+ \mathbf{a}_{||}+\beta \mathbf{a}_{\bot}}{1+ <\mathbf{a}|\mathbf{g}>/a_M^2},
\end{equation}
with $\beta=\sqrt{1-|\mathbf{g}|^2/a_M^2}$,
$\mathbf{a}_{||}=P_{\mathbf{g}}\mathbf{a}$ and
$\mathbf{a}_{\bot}=(I-P_{\mathbf{g}})\mathbf{a}$. This map defines
a projective symmetry \index{bounded symmetric domain!$D_a$}making
the ball $D_a$ into a \textit{bounded symmetric domain} with
respect to $Aut_p(D_a)$.

The existence of a maximal acceleration follows also from Born's
reciprocity principle, which states \cite{B65} that the laws of
nature are symmetric with respect to space and momentum. We think
that ``momentum" should be replaced by ``proper velocity" in
Born's reciprocity principle. Such reciprocity could not be
achieved with the Galilean  transformation for the proper-velocity
time continuum. As we have shown, this implies the existence of a
maximal acceleration. Caianiello's model \cite{Caianiello} also
supports Born's reciprocity principle.As shown by Schuller
\cite{shuller02}, the Muon Storage Ring
experiment\index{acceleration!maximal}
 (see \cite{Baily77} and \cite{Eisle87}) implies  that the constant
$\kappa<1.6\cdot 10^{-39}(s^2/m)^2$ and from the relativistic
correction of the Thomas precession $\kappa< 10^{-44}(s^2/m)^2$,
which is indeed close to zero. From Caianiello's model, the
estimate of the maximal acceleration in Scarpetta
\cite{scarpetta84} is $\kappa=0.4\cdot 10^{-103}(s^2/m)^2.$ But
the existence and the real value of maximal acceleration could be
obtained only from experiment, directly or indirectly.

\section{Space-time transformations to a uniformly accelerated
system without the Clock Hypothesis}

As we have shown, if the Clock Hypothesis is not valid, there
exist a unique maximal acceleration. Then from (\ref{tetap}) and
(\ref{1'p}), it follows that
\begin{equation}\label{mclockhyp}
  \tilde{t}=\sqrt{1-\frac{|\mathbf{g}|^2}{a_M^2}}t.
\end{equation}
This equation was called the \index{clock hypothesis!modified}
 \textit{modified Clock hypothesis}
by Shuller \cite{shuller02}. It shows that the dependence of the
rate of an accelerated clock is due to its \textit{acceleration}
as well as its velocity.

At this point we do not know how to obtain an explicit space-time
transformation from an inertial system $K$ to a system
$\widetilde{K}$ uniformly accelerated with respect to $K$ without
assuming the Clock hypothesis. But we can do this for any given
world-line $(t,\mathbf{r}(t))$ in $K$ by the following algorithm:

\textbf{Algorithm:} \textit{Step 1.} Use Lorentz transformations
to obtain a world-line $(t',\mathbf{r}'(t'))$ in $K'$ the comoving
frame to $\widetilde{K}$.

\textit{ Step 2.} Translate this world line $(t',\mathbf{r}'(t))$
from space-time representation to a world-line
$(t',\mathbf{u}'(t))$ in the p-velocity-time representation in
$K'$.

 \textit{Step 3.}  Use
transformation (\ref{cascadespacetimr1}) to obtain world-line
$(\tilde{t},\tilde{\mathbf{u}}(t))$ in the p-velocity-time
representation in $\tilde{K}.$ Finally,

 \textit{Step 4.} Using the  formula
  \begin{equation}\label{pvelspacetrorigin1}
 \mathbf{r}(t)=\int_0^{t} \mathbf{v}({\lambda})d\lambda =
  \int_0^{t}\frac{\mathbf{u}(\lambda)d\lambda}{\sqrt{1+|\mathbf{u}(\lambda)|^2/c^2}}
 \end{equation}
 obtain a  world-line $(\tilde{t},\tilde{\mathbf{r}}(t))$ in the
space-time representation in $\tilde{K}$ corresponding to the
world-line $(t,\mathbf{r}(t))$ in $K$.

We want to thank Michael Danziger for helpful remarks.

\end{document}